\documentclass[twocolumn]{aastex63}
\usepackage{natbib}
\graphicspath{{./}{figures/}}

\received{2021 May 26}
\revised{2021 June 17}
\revised{2021 June 25}
\accepted{}
\submitjournal{ApJ}

\shortauthors{Hasegawa et al.}

\begin{document}

\title{Discovery of two TNO-like bodies in the asteroid belt}

\correspondingauthor{Sunao Hasegawa}
\email{hasehase@isas.jaxa.jp}

\author[0000-0001-6366-2608]{Sunao Hasegawa}
\affiliation{Institute of Space and Astronautical Science, Japan Aerospace Exploration Agency, 3-1-1 Yoshinodai, Chuo-ku, Sagamihara, Kanagawa 252-5210, Japan}

\author[0000-0001-8617-2425]{Micha\"{e}l Marsset}
\affiliation{Department of Earth, Atmospheric and Planetary Sciences, MIT, 77 Massachusetts Avenue, Cambridge, MA 02139, USA}

\author[0000-0002-8397-4219]{Francesca E. DeMeo}
\affiliation{Department of Earth, Atmospheric and Planetary Sciences, MIT, 77 Massachusetts Avenue, Cambridge, MA 02139, USA}

\author[0000-0003-4191-6536]{Schelte J. Bus}
\affiliation{Institute for Astronomy, University of Hawaii, 2860 Woodlawn Drive, Honolulu, HI 96822-1839, USA}

\author[0000-0002-3291-4056]{Jooyeon Geem}
\affiliation{Department of Physics and Astronomy, Seoul National University, Gwanak-gu, Seoul 08826, South Korea}

\author[0000-0002-7332-2479]{Masateru Ishiguro}
\affiliation{Department of Physics and Astronomy, Seoul National University, Gwanak-gu, Seoul 08826, South Korea}
\affiliation{SNU Astronomy Research Center, Seoul National University, Gwanak-gu, Seoul 08826, Republic of Korea}

\author[0000-0002-8537-6714]{Myungshin Im}
\affiliation{Department of Physics and Astronomy, Seoul National University, Gwanak-gu, Seoul 08826, South Korea}
\affiliation{SNU Astronomy Research Center, Seoul National University, Gwanak-gu, Seoul 08826, Republic of Korea}

\author[0000-0002-7363-187X]{Daisuke Kuroda}
\affiliation{Okayama Observatory, Kyoto University, 3037-5 Honjo, Kamogata-cho, Asakuchi, Okayama 719-0232, Japan}

\author[0000-0002-2564-6743]{Pierre Vernazza}
\affiliation{Aix Marseille Univ, CNRS, LAM, Laboratoire d’Astrophysique de Marseille, Marseille, France}



\begin{abstract} 
Two extremely red main-belt asteroids: 203 Pompeja and 269 Justitia, were identified from combined visible and near-infrared spectroscopic observations collected at the IRTF and SAO observatories. 
These two asteroids have a redder spectral slope than any other D-type body, which are the reddest objects in the asteroid belt, and similar to RR and IR-class objects found in the outer Solar System among trans-Neptunian objects and Centaurs. 
Spectroscopic results suggest the presence of complex organic materials on the surface layer of these asteroids, implying that they could have formed in the vicinity of Neptune and been transplanted to the main belt region during a phase of planetary migration.
203 Pompeja is the only very red asteroid known so far among the $\sim$250 bodies with diameter larger than 110 km (i.e. presumably structurally intact) found in the asteroid belt.
These discoveries add another piece of evidence that the main asteroid belt hosts a population of bodies that were formed in the outskirt of the Solar System.
\end{abstract}

\keywords{Small Solar System bodies(1469) --- Asteroids (79) --- Trans-Neptunian objects(1705) --- Solar system formation(1530)}


\section{Introduction} \label{sec:intro}
Physical information about asteroids in the main-belt at the time when substantial radial mixing of planetesimals occurred is important to constrain the conditions of the formation of the Solar System.
Even after the Solar System entered a steady state at the end of the phase of planetary migration, the size distribution of main-belt asteroids (MBAs) continued to evolve due to collisional process. 
Although some D $\gtrsim$ 110km MBAs have been found to have fragmented through large collisions and subsequently reaccumulated \citep[e.g., ][]{Vernazza2020}, most MBAs larger than 110 km in diameter (D $\gtrsim$ 110 km MBAs) did not experience catastrophic destruction \citep[e.g., ][]{Bottke2005b}, meaning that these objects can be regarded as the last remains of the original population of planetesimals that initially populated the inner Solar System.
The fact that the majority of D $\gtrsim$ 110 km MBAs escaped catastrophic destruction also means that their orbital elements have not been substantially altered by collisions.
It is also known that orbits of D $\gtrsim$ 110 km MBAs are hardly changed by the Yarkovsky effect \citep[e.g., ][]{Vokrouhlicky2015}.
The fact that those orbital elements maintained their state at the end of the migration stage of the Solar System is important to constrain the conditions of formation of the Solar System.

The preserved information about the early Solar System contained in the physical and orbital properties of large (D $\gtrsim$ 110 km) MBAs motivated a survey of visible and near-infrared spectroscopic observations of these objects that aims to provide a better understanding of the original population of planetesimals that existed in the early Solar System.
In the course of this spectroscopic survey, we serendipitously discovered a very red spectral slope\footnote{A red spectral slope is defined as an increasing reflectance with increasing wavelength.} asteroid: 203 Pompeja with much redder spectral slope in the visible to near-infrared wavelength range than typical D-type MBAs and red Jovian Trojans.

Following this discovery, we searched in the literature for asteroids spectrally similar to 203, without limiting ourselves to diameters larger than 110 km, and retrieved a second very red asteroid: 269 Justitia.
In this paper, we present a comparison of these two very red asteroids with other Solar System bodies, and propose an interpretation of their composition and origin.

\section{Observations and Data Analysis} \label{sec:Observations and Data Analysis}
The new observations presented in this paper were performed on the 3.0-m NASA Infra-red Telescope Facility (IRTF) on Mt. Mauna Kea, Hawaii, USA (Minor Planet Center (MPC) code: 568) and at the 1.0-m Seoul National University Astronomical Observatory (SAO), Republic of Korea (no MPC code; 126\degr57\arcmin16\arcsec E, 37\degr27\arcmin25.7\arcsec N; 175 m) \citep{Im2021}. 
Asteroidal spectroscopic data were recorded by two different instruments mounted on these telescopes: the SpeX \citep{Rayner2003} installed at the f/35 Cassegrain focus of IRTF, and the PF0021VIS$-$LISA spectrometer attached to the f/6 Nasmyth focus of the SAO. 

Details about the observation conditions can be found in Appendix \ref{sec:203}.
More detailed information on the observing and reduction procedures can be found in \cite{DeMeo2009}, \cite{Marsset2020} for the IRTF data, and  \cite{Hasegawa2021} for SAO data.

\section{spectroscopic results} \label{sec:result}
The very red asteroid that we discovered by chance in the visible to near-infrared wavelength range during the spectroscopic survey for D $\gtrsim$ 110 km MBAs is 203 Pompeja, with a diameter of 111.3 km and an albedo of 0.045 (mean values obtained from IRAS \citep{Tedesco2002}, AKARI \citep{Usui2011}, and WISE \citep{Masiero2011}). 
This asteroid is located in the middle main-belt (\textit{a} = 2.737417 [au], \textit{e} = 0.058882, \textit{i} = 3.172125 [\degr])\footnote{The orbital elements of these asteroids were obtained from \url{https://ssd.jpl.nasa.gov/sbdb.cgi}\label{JPL}}.
The second very red asteroid in the visible to near-infrared wavelength range that had already been discovered \citep{DeMeo2009} is 269 Justitia, with a diameter of 54.4 km and an albedo of 0.080 (mean values obtained from IRAS \citep{Tedesco2002}, AKARI \citep{Usui2011}, and WISE \citep{Masiero2011}).
It is also located in the middle main-belt (\textit{a} = 2.616861 [au], \textit{e} = 0.213013, \textit{i} = 5.477865 [\degr])\textsuperscript{\ref{JPL}}.

\begin{figure*}
\gridline{\fig{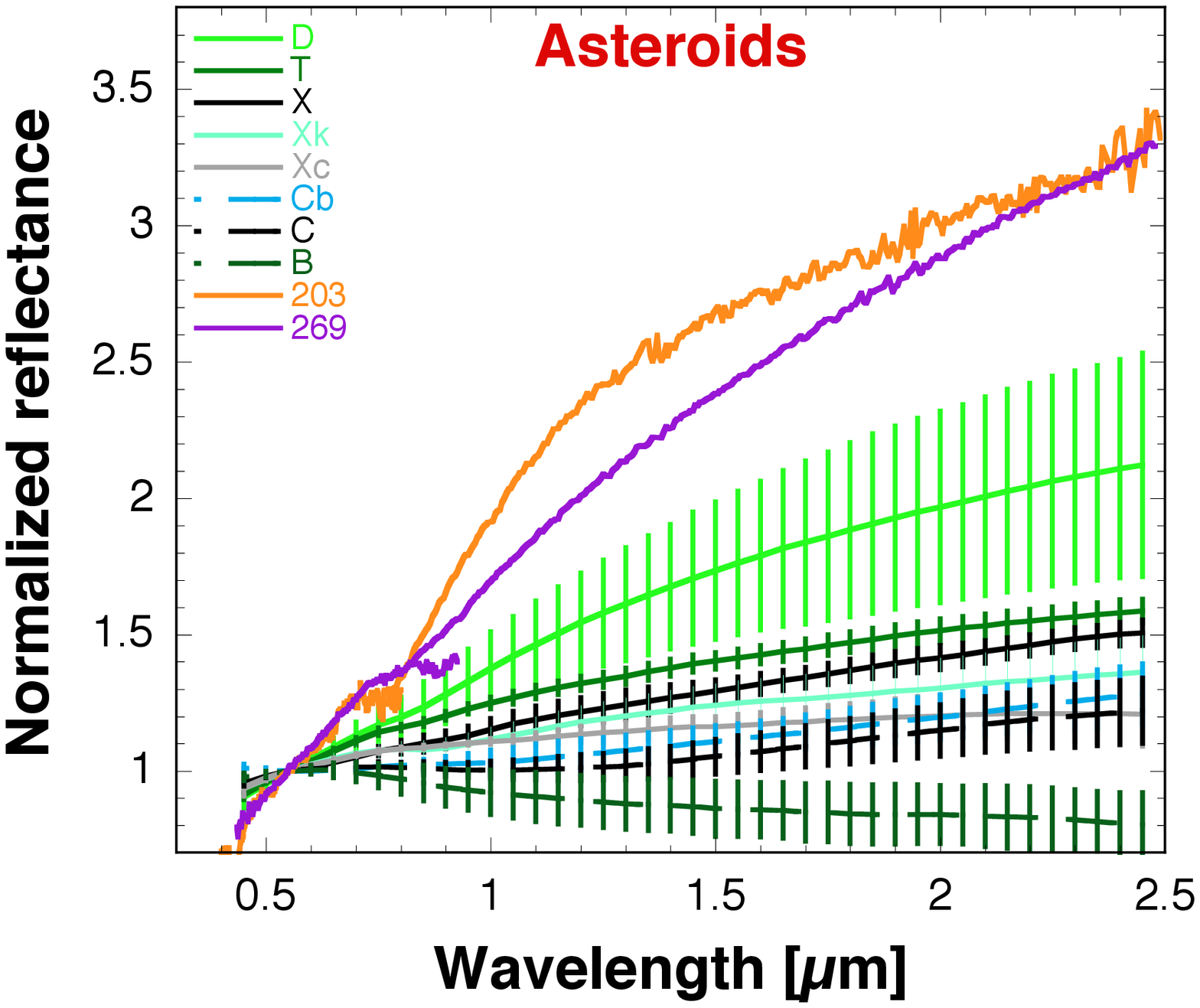}{0.5\textwidth}{}
          \fig{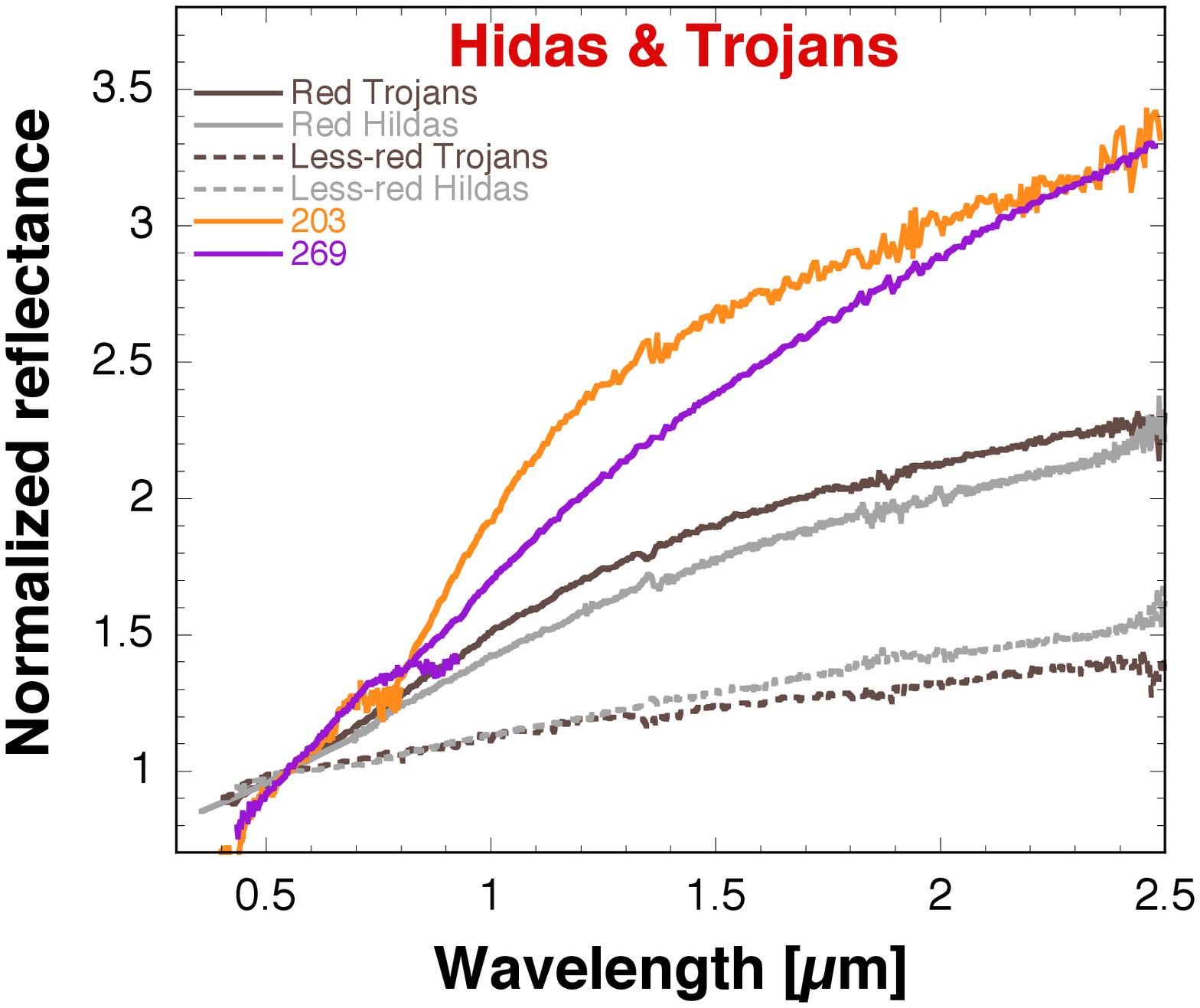}{0.5\textwidth}{}
          }
\gridline{\fig{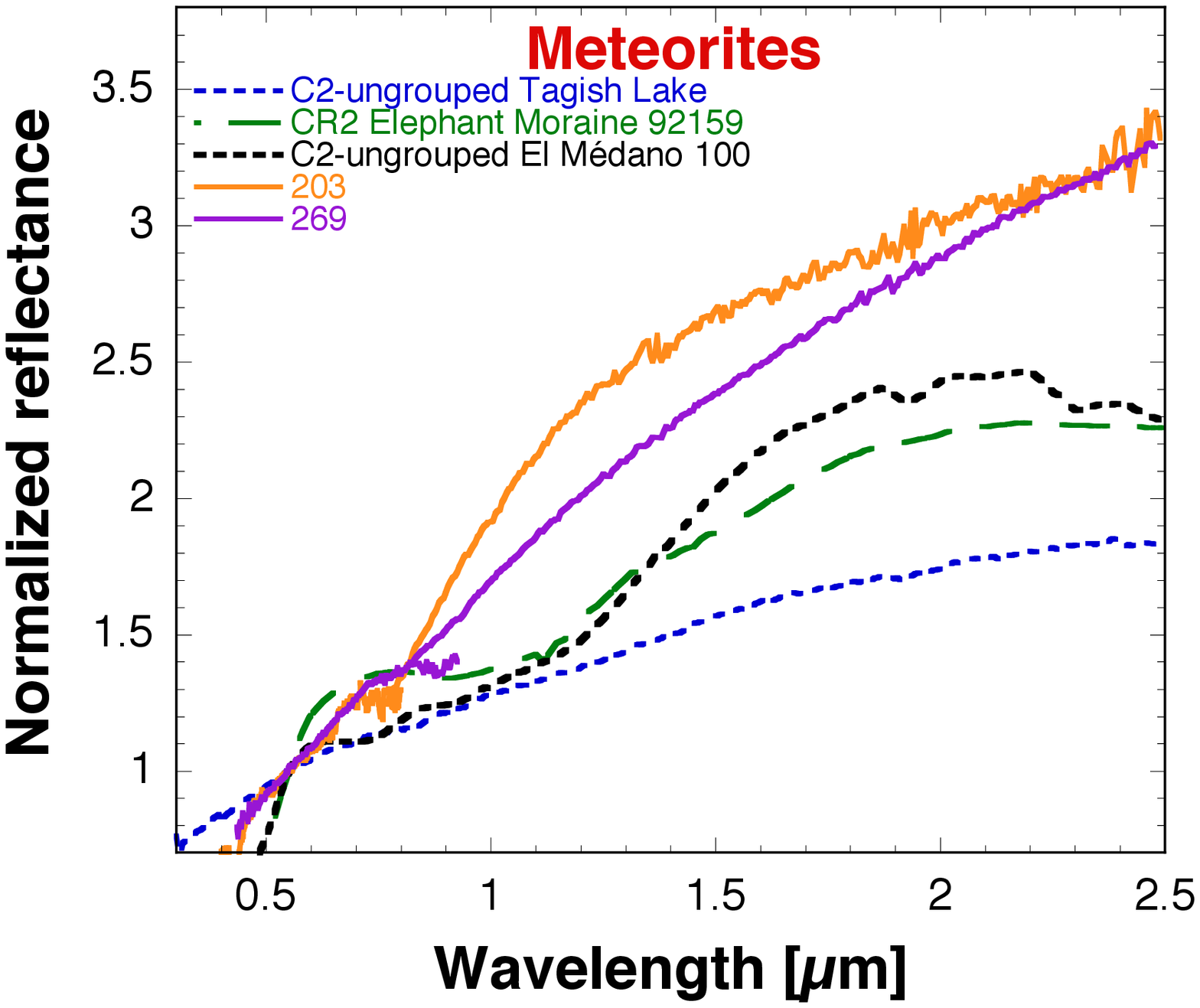}{0.5\textwidth}{}
          \fig{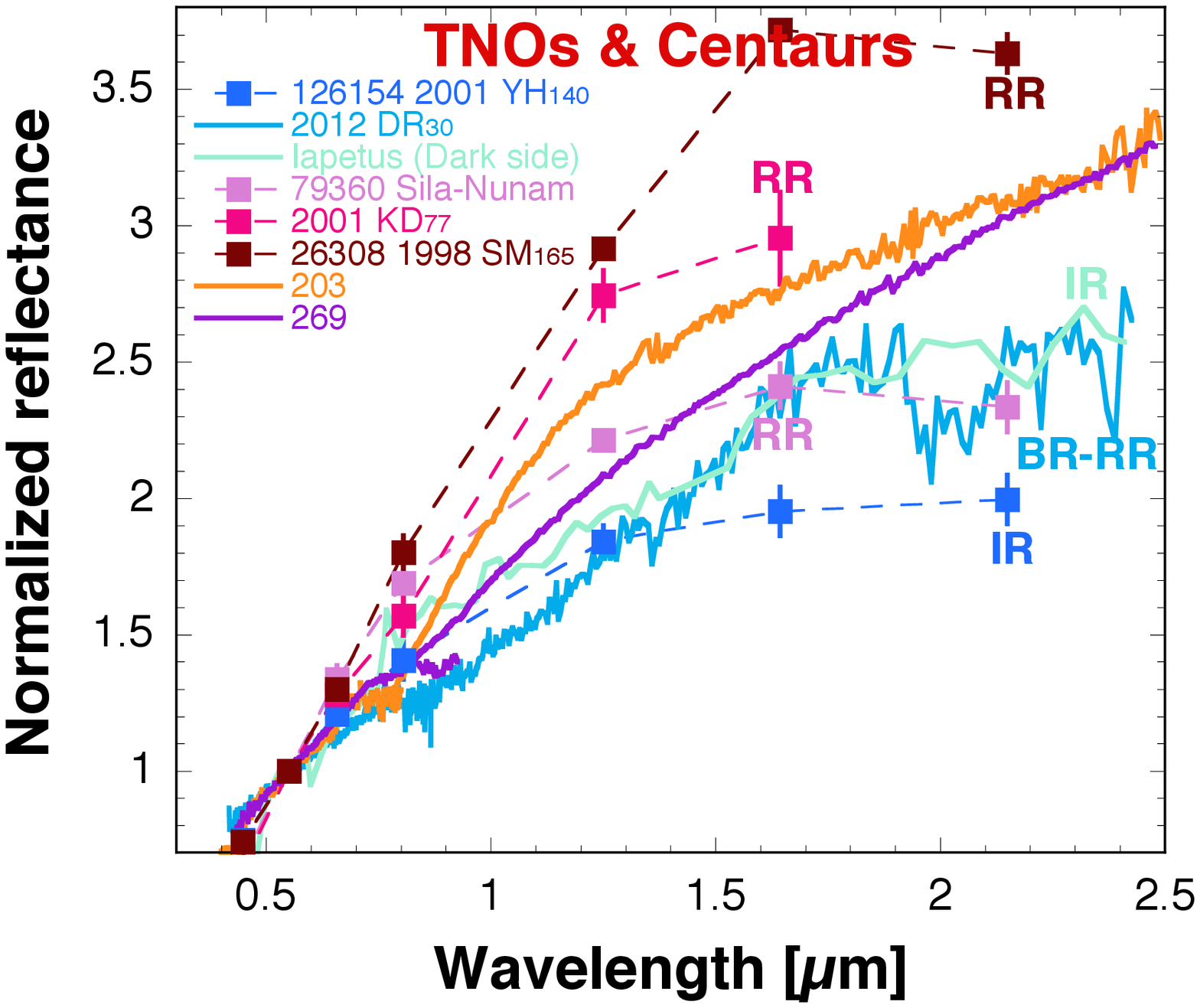}{0.5\textwidth}{}
          }
\caption{
Spectra of the very red spcetral slope MBAs 203 Pompeja and 269 Justitia and other dark (low-albedo) objects in the visible and near-infrared region.
Comparison of the very red asteroids with: \textit{top left} panel: typical spectral types of dark asteroids from the Bus-DeMeo classification scheme \citep{DeMeo2009}, \textit{top right}:  typical Hildas and Jovian Trojans \citep{Emery2011, Wong2017}, \textit{bottom left}: meteorites with a very red spectral slope  \citep{Hiroi2001, Beck2018, KrmerRuggiu2021}, \textit{bottom right}: dark outer Solar System objects  spectrally similar to the very red asteroids \citep{Cruikshank1987, Fernandez-Valenzuela2021, Hainaut2012, Szabo2018}
}
\label{fig:spectra}
\end{figure*}

\begin{figure*}
\gridline{\fig{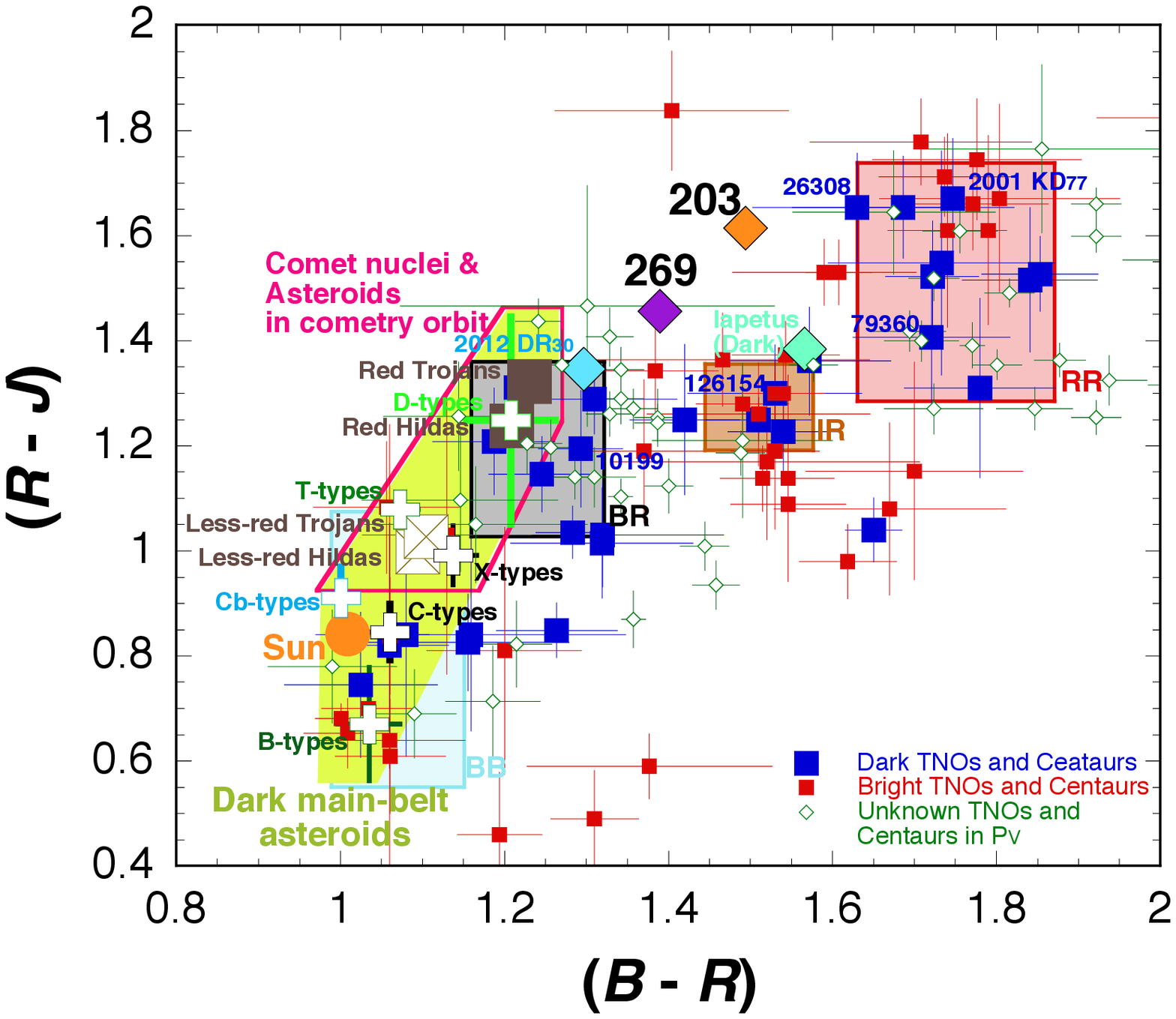}{0.45\textwidth}{}
          \fig{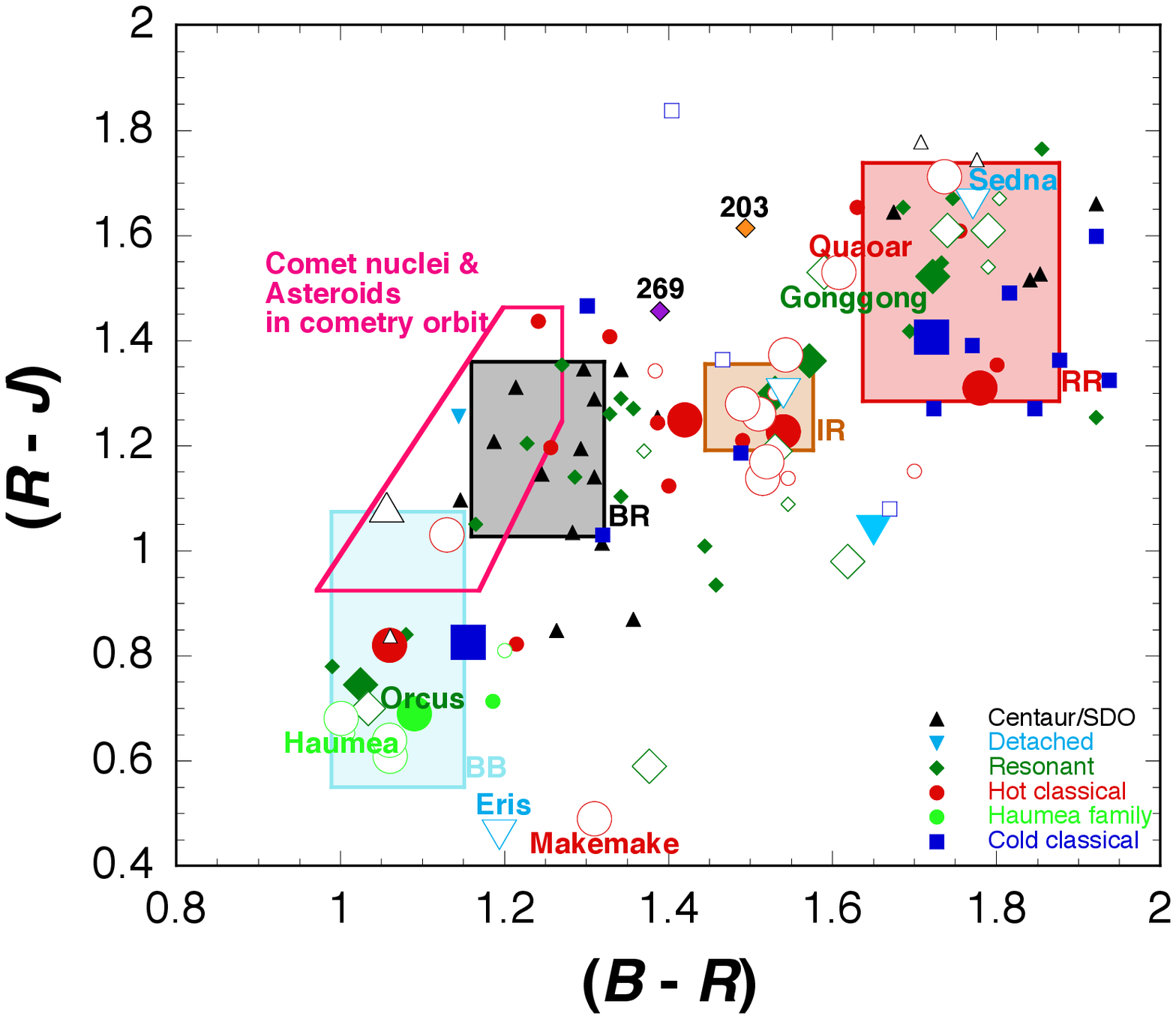}{0.45\textwidth}{}
          }
\caption{(\textit{B} $-$ \textit{R}) versus (\textit{R} $-$ \textit{J}) color$-$color plot of airless objects. 
Color data of typical dark spectral-type asteroids, Hildas, and Jovian Trojans from \cite{DeMeo2009}, \cite{Emery2011}, and \cite{Wong2017}, respectively.
Photometric data for TNOs and other icy bodies are taken from \cite{Hainaut2012}, \cite{Fernandez-Valenzuela2021}, \cite{Schwamb2019}, \cite{Szabo2018}, \cite{Cruikshank1987}, and \cite{Perna2010}.
Colorimetric data of the comets, asteroids in cometary orbit, and interstellar objects are adapted from Table \ref{tab:B1} in Appendix \ref{sec:listB1}.
Albedo's data is cited from the following: \cite{Mueller2020}, \cite{Thomas2000}, \cite{Tedesco2002}, \cite{Masiero2011}, and \cite{Usui2011}.
Left: Figure showing the classification in terms of albedo.
Right: Figure showing the classification in terms of dynamical classes.
Large and open symbols are objects with diameters larger than ${10^{2.5}}$ km, that almost never have BR-type spectra (also see Figure \ref{fig:SizeDis}),  and bright objects whose albedo is known to be greater than 0.1, respectively.
The sizes of objects for which the albedo is not known are estimated by assuming an albedo of 0.1.
\label{fig:BRvsRJ-pv}
}
\end{figure*}

Since the discovered very red asteroids are dark asteroids with albedo less than 0.1, a comparison of their spectra with those of typical dark MBAs in the visible to near-infrared wavelength range \citep{DeMeo2009} was carried out (Figure \ref{fig:spectra}).
The results show that the two very red asteroids exhibit very red spectral slopes in the near-infrared wavelength range compared to typical D-type bodies, which constitute the reddest bodies in the Bus-DeMeo classification of asteroids \citep{DeMeo2009}.
For a quantitative comparison, we made a (\textit{B} $-$ \textit{R}) versus (\textit{R} $-$ \textit{J}) color$-$color plot (Figure \ref{fig:BRvsRJ-pv}).
Typical D-type asteroids have (\textit{B} $-$ \textit{R}) and (\textit{R} $-$ \textit{J}) values of 1.21 and 1.25, whereas values of (\textit{B} $-$ \textit{R}) and (\textit{R} $-$ \textit{J}) for 203 and 269 are 1.49 and 1.61, and 1.39 and 1.46, respectively.
Like D-type MBAs, the very red asteroids exhibit no strong absorption feature in their spectra. 
It is further known that asteroids in the Hilda and Jovian Trojan groups have spectra similar to those of D-type asteroids \citep{DeMeo2014b}.
We find that the spectra of the very red asteroids have redder spectral slopes than those of the Trojan and Hilda groups \citep{Emery2011, Wong2017}  in the near-infrared wavelength range (Figure \ref{fig:spectra}).
The values of (\textit{B} $-$ \textit{R}) and (\textit{R} $-$ \textit{J}) for red Hildas and Jovian Trojans are 1.21 and 1.24, and 1.23 and 1.32, respectively.
In short, 203 Pompeja and 269 Justitia have a very red slope spectrum in the visible and near-infrared wavelength range compared to a wide range of inner Solar System objects, ranging from the asteroid belt down to the near-Earth asteroids out to the Jovian Trojans which are typically classified as ``asteroids''.

We further searched for spectra of meteorites with very red spectral slope in the literature and identified three objects that we compared to the very red asteroids (Figure \ref{fig:spectra}): The C2-ungrouped El M\'{e}dano 100 \citep{KrmerRuggiu2021}, the CR2 Elephant Moraine 92159 \citep{Beck2018}, and the C2-ungrouped Tagish Lake \citep{Hiroi2001} meteorites.
The spectra of the first two meteorites are characterized by a deep 1-\micron\ absorption feature that differentiate them from the spectra of 203 and 269, and make them similar to the spectra of A-type asteroids, whose surfaces are thought to be made of olivine reddened by space weathering. 
The spectrum of the Tagish Lake meteorite, on the other hand, which is thought to have originated from some D-type asteroid \citep{Hiroi2001}, provides a better spectral match to 203 and 269, but its spectral slope is far from less red. 
Since no meteorite with a redder D-type asteroid-like spectrum than the Tagish Lake meteorite has ever been found, we conclude that no meteorite originating from a very red asteroid has been found so far.

It is known that some trans-Neptunian objects (TNOs) and Centaurs have a very red spectrum in the visible wavelength range compared to Jovian Trojans \citep[e.g., ][]{Wong2016}.
We compared the spectra of the very red asteroids to spectra of outer Solar System objects (TNOs and Centaurs) categorized in the typical classification scheme for TNOs \citep{Fulchignoni2008, Perna2010}\footnote{In this classification scheme, four broad classes of objects, named BB, BR, IR and RR, were defined. BB objects have neutral or blue spectrum.  RR objects exhibit a very red sloping spectrum. BR and IR objects have spectra with intermediate inclinations between BB and RR objects, with BR having a bluer spectrum.}.
We acknowledge, however, that alternative classification systems of TNOs have since been proposed \citep[e.g., ][]{Schwamb2019}.
In the visible light range, the surface of the very red asteroids are less red than the surface of the IR objects, whereas in the full visible to near-infrared light range, the very red asteroids are as red as IR objects and RR objects (Figure \ref{fig:spectra}).
Values of the normalized reflectance in the \textit{K}-band for RR and IR objects are 2.6 and 2.0, respectively, while they are 3.1 and 3.0 for 203 Pompeja and 269 Justitia, respectively, i.e., rather steeper than the typical RR and IR objects.
The spectra of very red asteroids fall within the range of spectra of TNOs and Centaurs (Figure \ref{fig:spectra}).
We further note that the polarization properties of 269 retrieved from the literature reveal additional similarities to that of some icy solar system bodies and, in particular F-type bodies, while dark (low-albedo) other C-complex and Trojan asteroids provide poorer fits (see Appendix \ref{sec:result2}).

The spectra of dark objects with colors similar to 203 Pompeja and 269 Justitia in Figure \ref{fig:BRvsRJ-pv} are shown in the third panel of Figure \ref{fig:spectra}.
It has been pointed out that complex organic materials such as Tholin \citep[e.g., ][]{Sagan1979} exist on the surface of these bodies whose names are listed in the legend of Figure \ref{fig:spectra} \citep{DalleOre2013, Szabo2018, Owen2001}.
This indicates that the surfaces of 203 Pompeja and 269 Justitia could also have complex organic materials to explain their spectral similarities to these objects.
On the other hand, it has been reported in \cite{Brunetto2006} that simple organic materials irradiated with ions form refractory residues with a very red slope spectrum.
It is possible that the red spectral slope spectra of 203 and 269 resulted from a similar process.

The visible ($\lambda$ $<$ 0.8 $\mu$m) slopes of 203 and 269 are not as red as those of RR TNOs (Figure \ref{fig:BRvsRJ-pv}) and their albedos are darker \citep[e.g., ][]{Hasegawa2017}.
\cite{Grundy2009} showed that the visible slope and albedo for mixtures of volatile ice and non-volatile organics in the visible wavelength range become less red and less bright as water ice becomes less abundant.
The less-red visible slope, darker albedos and continuous positive near-infrared slope up to 2.5 \micron\ of 203 and 269 therefore suggest that their surfaces are depleted in water with respect to RR TNOs.
The story, however, might be more complicated, as telescopic observations from \cite{Barkume2008} found little correlation between the percentage of water ice in TNOs and Centaurs, and the optical colors and albedos of these objects. 
\cite{Merlin2017}, on the other hand, confirmed the spectroscopic trend found by \cite{Barucci2008} on the increasing occurrence of water within taxonomical classes, from IR/BR to RR and BB. 
As shown in their paper, some of the TNOs and Centaurs show absorption features near the \textit{H}- and \textit{K}-bands attributable to water, methanol, and hydrocarbons ices, while 203 and 269 show no absorption at all in these bands. 
This confirms a lack of volatile ices in the surface layer of these two very red asteroids. 
It is also possible that increased space weathering at the heliocentric distance of the very red asteroids preferentially removed absorption bands from the spectra of these objects compared to Centaurs and TNOs \citep[e.g., ][]{Brunetto2006}.

\section{Discussion} \label{sec:Discussion}
Our spectroscopic analysis suggests the presence of a mixture of complex organics and other unknown and spectrally featureless materials (such as, for instance, small grained and/or iron-poor silicates) on the surface of the very red asteroids 203 Pompeja and 269 Justitia.
We exclude the presence of water ice at their surface owing to their relative proximity to the Sun.
\cite{Jewitt2015} suggested that complex organics is covered by the onset of cometary activity or ejected upon entry into the inner Solar System to explain the lack of red materials inside the heliocentric distance at which Centaurs and comets become active (about 10 au).
In order for the red spectral slope material to be available in the surface layers of the very red asteroids, while being absent in comets and active Centaurs, there must be an insulating blanket on their surface thick enough so that cometary activity did not eject or cover the layer of red material.
However, the typical size of a comet is considered to be about 10 km \cite[e.g., ][]{Stern2003}, which is about 0.5 to 1 order of magnitude smaller than the diameter of 203 and 269.
It is generally thought that different sizes of minor bodies result in different evolution of the surface due to the thermal evolution of the interior \cite[e.g., ][]{McSween2002, Guilbert-Lepoutre2020}.
To test that hypothesis, we further investigate composition change with size for TNOs and Centaurs.

\begin{figure*}
\gridline{\fig{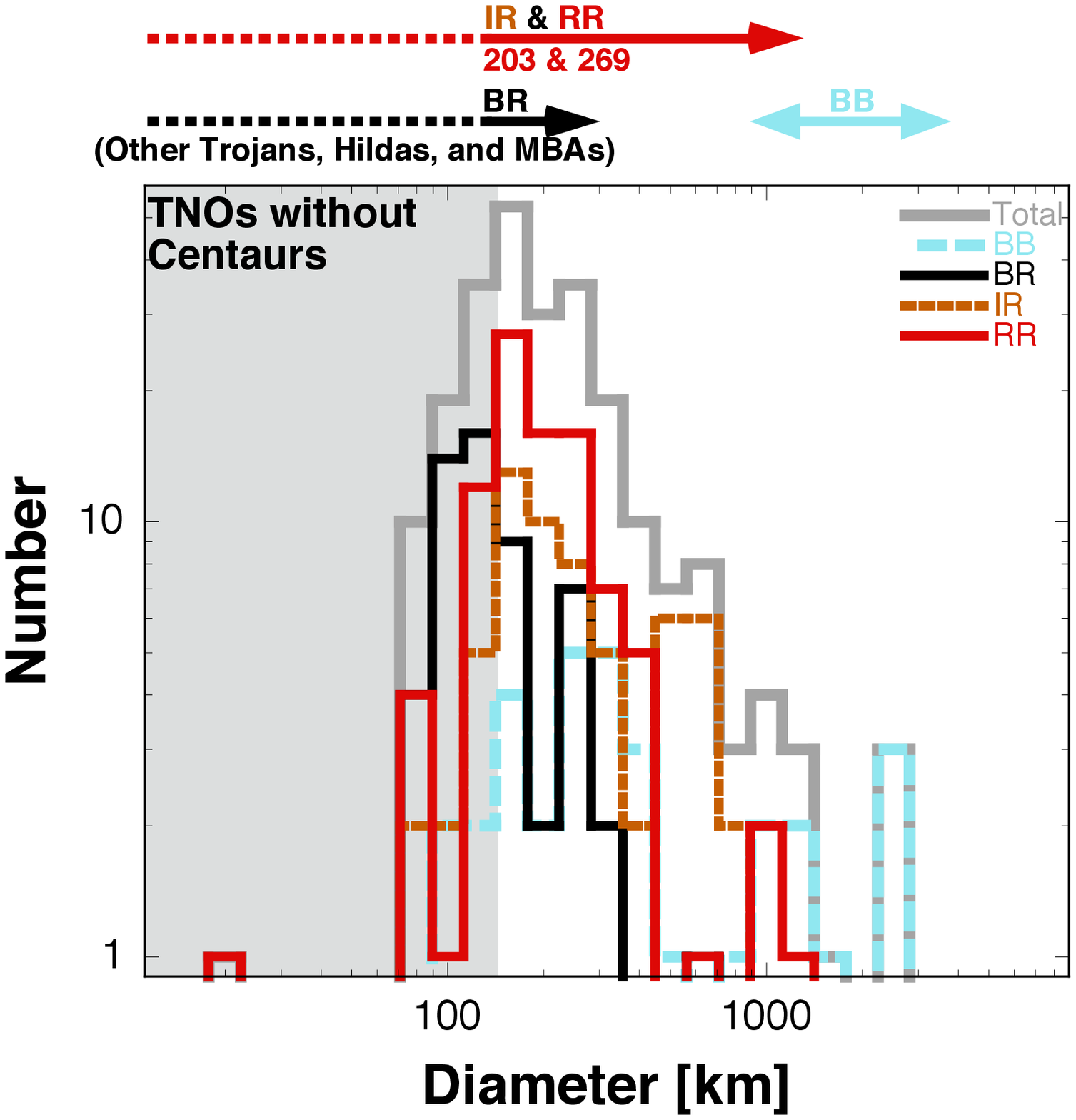}{0.5\textwidth}{}
          \fig{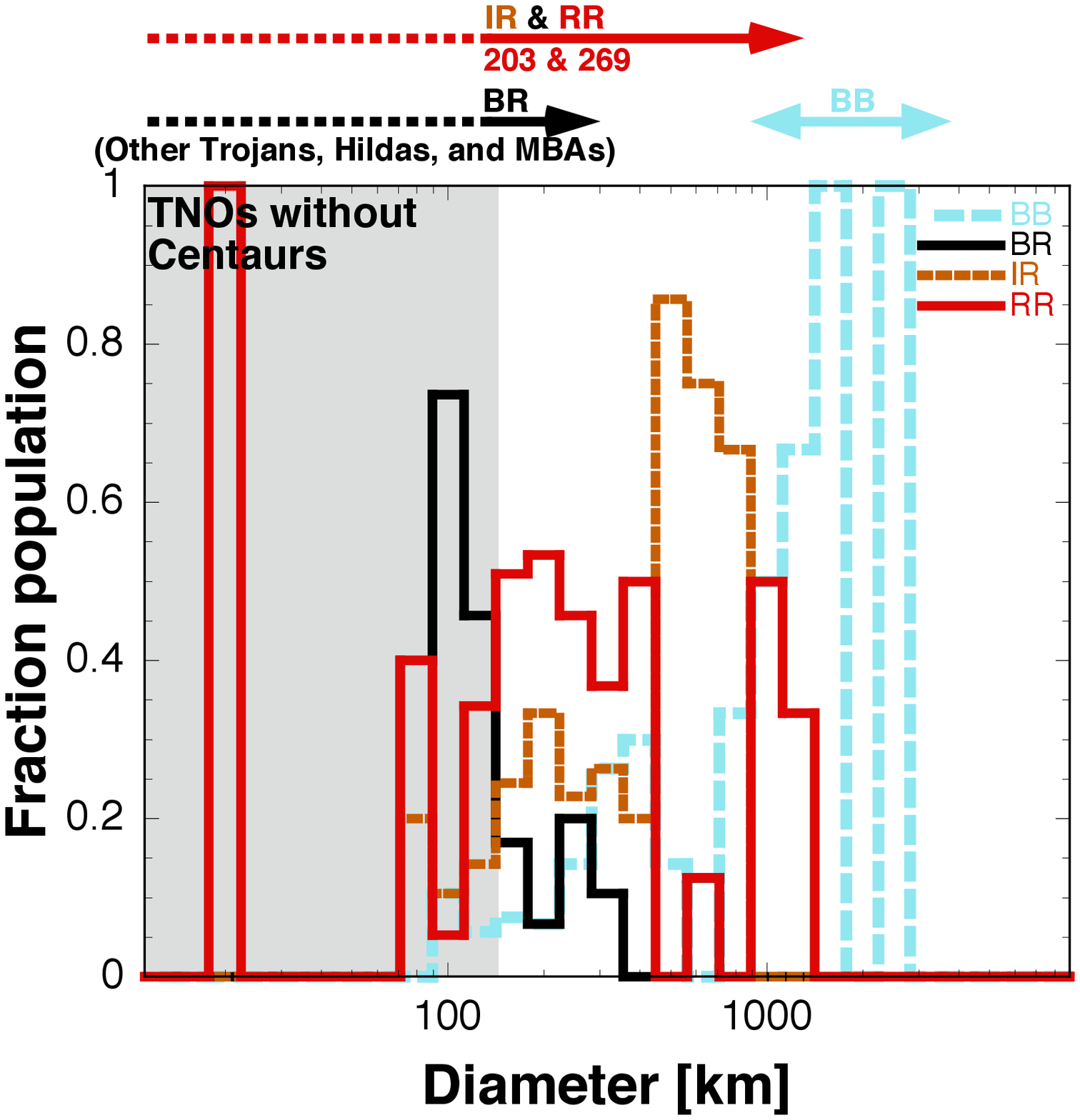}{0.5\textwidth}{}
          }
\caption{
Histogram of the size distribution of TNOs without Centaurs classified following the TNO spectral classification method of \cite{Fulchignoni2008}.
In addition to the data used in Figure \ref{fig:BRvsRJ-pv}, data of TNO that do not include the near-infrared wavelength region \citep{Belskaya2015} include Triton \citep{Hicks2004, Grundy2010}, Charon \citep{Protopapa2020}, and 486958 Arrokoth \citep{Grundy2020}.
When the size of the object was not known, it was obtained by assuming an albedo of 0.1.
Unclassified TNOs were classified through the color$-$color diagram in Figure \ref{fig:BRvsRJ-pv}.
Looking at the distribution of the number of TNOs, we note that their number decreases below ${10^{2.1}}$ km due to the incompleteness size limit of the survey.
Hence, the values in the gray shaded region are not used in the discussion.
\label{fig:SizeDis}
}
\end{figure*}

Figure \ref{fig:SizeDis} shows the histogram of sizes of each spectral type of Centaurs and TNOs, including Neptune I Triton, 134340 Pluto, and Pluto I Charon. 
We include both measurements covering the full visible and infrared wavelength region, as well as measurements taken in visible wavelengths only.
TNOs with diameters larger than about 1000 km are dominated by BB-type (neutral or blue slope) objects, but the fraction of BB-type objects decreases as diameter decreases below about 1000 km.
\cite{Guilbert-Lepoutre2020}  showed that TNOs  larger than about 500 km in diameter could have been strongly altered by geological activity that modified both their internal structure and their surface.
In addition, \cite{Brown2007} showed that the Haumea family was formed by a single catastrophic impact and that the family members are fragments of the ice mantle of 136108 Haumea.
From the aforementioned two hypotheses, it is thought that the red material formed on the surface of the BB-type objects partially or entirely disappeared due to geological activity or disruptive impacts.
Both RR (very red slope) and IR (moderate red slope) TNOs are smaller than about 1500 km, with the $\sim$800 to $\sim$200 km range accounting for about 70 \% of the total of RR and IR TNOs.
The increase in number/fraction of both the RR- and IR-types TNOs below about 1500 km may be due to the lack or decline of refreshment of their surface layer as geological activity decreases as a function of size \citep{Guilbert-Lepoutre2020}.
The Voyager 2 and New Horizons mission revealed that Triton and Pluto, which are both larger than 1500 km in diameter, exhibit geologically young surfaces that appear to have been active until recently due to cryovolcamism and tectonic structures, respectively \citep{Smith1989, Moore2016}.
It is possible that the remaining BB objects with diameters of less than about 800 km are ejecta from large BB-type TNOs such as Haumea.
However, out of the 32 BB bodies that we retrieved from the literature, 10 only are members of the Haumea family, implying that Haumea is not the only source of small BB TNOs. 
Considering that the Haumea family is probably the only collisional family in the TNO region \citep{Levison2008}, it is possible that the remaining small BB TNOs are the remaining fragments of large TNOs destroyed during the late phase of planetary migrations. 
According to \cite{Brunetto2006}, organic ice reddens in about 1 Gyr at most, resulting in a red spectral slope like the RR objects, so water ice that does not redden must be exposed on the surface layer on the such BB objects.
Since some comets and asteroids with cometary orbits have spectra of X-complex asteroids i.e. similar to BB-type objects (see Figure \ref{fig:BRvsRJ-pv} and Table \ref{tab:B1}), the origin of BB-type asteroids with diameters less than 300 km may be similar to that of D-type/BR-like asteroids with diameters less than 300 km (see next paragraph).

The BR-type TNOs, which have similar spectral properties to D-type asteroid and most comets and asteroids with cometary orbits (see Figure \ref{fig:BRvsRJ-pv} and  Table \ref{tab:B1}), appear below a diameter of about 300 km.
Since most comets and asteroids with cometary orbits, generally considered to be survivors of planetesimals, predominantly have D-type or X-complex spectra, it is possible that BR-type objects in outer Solar System, which are spectrally similar to D-type asteroids, are also leftover of a previous generation of planetesimals.
The maximum diameter of D-type asteroids in the main-belt and Jovian Trojans is about 200 km \citep{Vernazza-inpress}, which agrees with the existence of the BR-type TNOs with diameters less than 300 km.
Also, most BR types have albedo less than 0.1 (see Figure \ref{fig:BRvsRJ-pv}), which is consistent with the average albedo of D-type asteroids.
These results support that most Hildas and Jovian Trojans, as well as comets including asteroids with comet-like orbits, and BR TNOs, have a common composition and that most Hildas and Jovian Trojans are embedded objects from the trans-Neptunian planetesimal region.

On the other hand, 203 Pompeja and 269 Justitia retain spectra similar to the IR- and RR-types. 
Although Jovian Trojans and Hildas, and Cybeles which are positioned in the outer main-belt, are dominated by D $\gtrsim$ 110 km asteroids with a BR-type spectrum \citep[e.g., ][]{DeMeo2014b}, 203 Pompeja is the only very red MBA discovered so far out of about 250 MBAs with a diameter larger than 110 km.
The eccentricities of 203 Pompeja and 269 Justitia are much smaller than those of typical short-period comets and Centaurs, suggesting that these very red asteroids did not come to their present position through scattering interactions with Neptune after the migration stage of the early Solar System.
We therefore need a plausible explanation for why these two MBAs are still very red, while most of the remaining Hildas and Jovian Trojans are less red, and why D-types are much more abundant than very red asteroids such as 203 and 269.

It is generally thought that dark C- and X-complexes and T, and D-types accreted beyond the water ice snowline like TNOs \citep[e.g., ][]{Vernazza2017}.
However, in order to create a surface layer with an extreme red (RR) spectrum, it is necessary to have complex organic matter.
In order to produce complex organic matter, the presence of volatile substances other than water ice, such as ices made of volatile organic matter (e.g., methanol and methane), is necessary \citep[e.g., ][]{Sagan1979}.
Therefore, IR and RR objects are likely to have formed beyond the snowline of some volatile organics, while BR objects would have formed inside \citep[e.g., ][]{Vernazza2017}.
In that scenario, the fact that the number of very red asteroids is much smaller than that of D-type asteroids (similar to BR-type TNOs) would simply mean that these objects came from farther away.
The implantation of outer solar system bodies in the asteroid belt during the early stage of solar system formation is supported by dynamical models \citep[e.g., ][]{Raymond-inpress}, as well as by the recent discovery of a ${\mathrm{CO_{2}}}$-bearing fluid in the Sutter's Mill carbonaceous chondrite, which suggest that this meteorite formed beyond the ${\mathrm{CO_{2}}}$ snowline \citep{Tsuchiyama2021}.
The fact that cold classical TNOs have a redder surface layer in the visible wavelength range than most dynamically excited TNOs, suggesting that the cold classical TNOs formed further away from the Sun than dynamically excited objects \citep[e.g., ][]{Schwamb2019}, is consistent with this hypothesis.

Assuming 203 Pompeja and 269 Justitia indeed migrated from the trans-Neptunian planetesimal region into the middle main-belt, as suggested by their similar spectral and polarimetric properties to outer Solar System objects, they must have retained some fraction of water ice in their blankets during the migration phase.
1 Ceres is considered to have water ice inside its outer crust \citep{McCord2018}, and if the thick blanket of Ceres, like the very red asteroids's blanket, is composed of organic matter, the surface layer must have avoided being stripped during its migration from the trans-Neptunian planetesimal region.
Based on \cite{Bottke2005b}, it is thought that 203 Pompeja, with a diameter of about 110 km, has not experienced a destructive impact, but 269 Justitia, because of it is diameter about 54 km, is thought to have experienced a catastrophic impact, so there may be an unknown mechanism that can retain the complex organic blanket even after a destructive impact.
Alternatively, maybe it just so happens that 269 Justitia also avoided any destructive impacts.

\section{Summary} \label{sec:summary}
We serendipitously discovered a very red asteroid: 203 Pompeja, while conducting a spectroscopic survey of D $\gtrsim$ 110 km MBAs.
We further retrieved another very red asteroid: 269 Justitia with a diameter of about 54 km in the past literature.

 \begin{enumerate}
\item The very red asteroids have less-red spectral slope than IR-type outer Solar System objects (TNOs and Centaurs) in the visible wavelength range. Meanwhile, their surfaces are as red as IR- and RR-type objects in the overall visible and near-infrared wavelength range.
\item The spectroscopic similarities of the very red asteroids to outer Solar System objects suggest that complex organic materials dominate their surface.
\item We therefore propose that the very red asteroids formed in the trans-Neptunian planetesimal region beyond the snowline of volatile organic materials that originally separated BR and IR/RR TNOs.
This is consistent with theoretical scenarios \citep[e.g., ][]{Raymond-inpress} which propose that some planetesimals formed as far away as 20-30 au could have been implanted in the main-belt.
\end{enumerate}

The discovery of these objects poses important constraints on the theory of Solar System formation and subsequent dynamical evolution.
\\
\\

\begin{deluxetable*}{rlcccl}
\tablenum{B1}
\tablecaption{List of the comets, asteroids in cometary orbit, and interstellar objects}
\tablewidth{0pt}
\tablehead{
\colhead{Num}&\colhead{Name}&\colhead{Orbit-type}&\colhead{Observation}&\colhead{Taxonomy}&\colhead{Reference data} 
}
\startdata
358P&PANSTARRS &Encke-type comet &Vis+NIR&X&\cite{Snodgrass2017}\\ 
28P&Neujmin&Jupiter-family comet&Vis+NIR&D&\cite{Campins2007}\\ 
66P&du Toit&Jupiter-family comet&Vis+NIR&D&\cite{Yang2019}\\ 
67P&Churyumov-Gerasimenko&Jupiter-family comet&Vis+NIR&D&\cite{Capaccioni2015}\\ 
162P&Siding Spring&Jupiter-family comet&Vis+NIR&D&\cite{Campins2006}\\ 
&C/2001 ${\mathrm{OG_{108}}}$&Halley-type comet&Vis+NIR&X&\cite{Abell2005}\\ 
&C/2013 ${\mathrm{P_{2}}}$&Comet not in category&Vis+NIR&BR-IR&\cite{Meech2014DPS}\\ 
944&Hidalgo&Asteroid in JFC$^{(a)}$&Vis+NIR&D&\cite{Licandro2018}\\ 
3552&Don Quixote&Asteroid in JFC$^{(a)}$&Vis+NIR&D&\cite{Licandro2018}\\ 
6144&Kondojiro&Asteroid in JFC$^{(a)}$&Vis+NIR&D&\cite{Licandro2018}\\ 
20898&Fountainhills&Asteroid in JFC$^{(a)}$&Vis+NIR&D&\cite{Licandro2018}\\ 
30512&2001 ${\mathrm{HO_{8}}}$&Asteroid in JFC$^{(a)}$&Vis+NIR&D&\cite{Licandro2018}\\ 
347449&2012 ${\mathrm{TW_{236}}}$&Asteroid in JFC$^{(a)}$&Vis+NIR&D&\cite{Licandro2018}\\ 
&2005 ${\mathrm{NA_{56}}}$&Asteroid in JFC$^{(a)}$&Vis+NIR&X&\cite{Licandro2018}\\ 
65407&2002 ${\mathrm{RP_{120}}}$&Damocloid$^{(b)}$&Vis+NIR&D&\cite{Licandro2018}\\ 
&1996 PW&Damocloid$^{(b)}$&Vis+NIR&D&\cite{Davies1998}\\ 
&1998 ${\mathrm{WU_{24}}}$&Damocloid$^{(b)}$&Vis+NIR&D&\cite{Davies2001}\\ 
1I&'Oumuamua&Interstellar asteroid&Vis+NIR&D&\cite{Fitzsimmons2018}\\ 
\enddata
\tablecomments{
$^{(a)}$Asteroid with the orbit of a Jupiter-family comet.
$^{(b)}$Asteroid with the orbit of a Halley-type or long-period comet.}

\label{tab:B1}
\end{deluxetable*}

We would like to thank the referee for the careful review and constructive suggestions, which helped us to improve the manuscript significantly. 
We are grateful to Dr. Ian Wong for sharing their valuable spectral data.
This work is based on observations collected at the Infrared Telescope Facility, which is operated by the University of Hawaii under contract 80HQTR19D0030 with the National Aeronautics and Space Administration and at Seoul National University Astronomical Observatory. 
The authors acknowledge the sacred nature of Maunakea and appreciate the opportunity to observe from the mountain. 
We are grateful to Dr. Tomohiko Sekiguchi, Mr. Koki Takahashi, Ms. Kana Makino, Dr. Tom Seccull, Mr. Sunho Jin, Mr. Yoonsoo P. Bach, and Mr. Jinguk Seo for their support.
M.M. and F.D. were supported by the National Aeronautics and Space Administration under grant No. 80NSSC18K0849 and 80NSSC18K1004 issued through the Planetary Astronomy Program.
M.Im acknowledges the support from the Korea Astronomy and Space Science Institute grant under the R\&D program (Project No.2020-1-600-05) supervised by the Ministry of Science and Technology and ICT (MSIT), and the National Research Foundation of Korea (NRF) grant, No. 2020R1A2C3011091, funded by MSIT. 
M.Is. was supported by the NRF grant No. 2018R1D1A1A09084105.
This study was supported by JSPS KAKENHI (grant nos. JP18K03723, JP19H00719, JP20K04055, JP21H01140, and JP21H01148) and by the Hypervelocity Impact Facility (former facility name: the Space Plasma Laboratory), ISAS, JAXA.

\appendix
\section{Observations of 203 Pompeja  and 269 Justitia} \label{sec:203}
\paragraph{203 Pompeja}
Visible and near-infrared spectroscopic observations of 203 Pompeja were made in 2021 Jan 29 and 2021 Jan 14, respectively.
Single exposure integration time and the number of spectral files used to produce the final combined spectrum of the observations in visible and near-infrared wavelength range are 300 sec $\times$ 4 and 120 sec  $\times$ 18, respectively.
Photometric observations of the asteroid in the \textit{V}-, \textit{R}-, and \textit{I} -bands were also made in 2021 Feb 4 to confirm the legitimacy of the visible spectroscopic observations (Figure \ref{fig:203-269}).

\paragraph{269 Justitia}
The spectrum of 269 Justitia in visible and near-infrared range was used from the data published in \cite{Bus2002} and \cite{DeMeo2009}.
However, it later was found that there was also unpublished data of 269, so that data was also used as well.
Single exposure integration time and the number of spectral files used to produce the final combined spectrum of the observation in the near-infrared wavelength range is 120 sec  $\times$ 14.

The intrinsic 1$\sigma$ slope uncertainty over 0.8 to 2.4 $\mu$m of spectrum by SpeX observations is 4.2 \% $\mu$m$^{-1}$ \citep{Marsset2020}.
The difference between the two spectra of 269 is about 2 \% at 2.5 $\mu$m, which is smaller than the uncertainty, implying that the two spectra match within the measurement error range.
In these two observations, 269 does not show any evidence for surface variability.

The spectrum of 269 Justitia in used this study is a simple average of the two spectra (Figure \ref{fig:203-269}).

\restartappendixnumbering
\begin{figure*}
\gridline{\fig{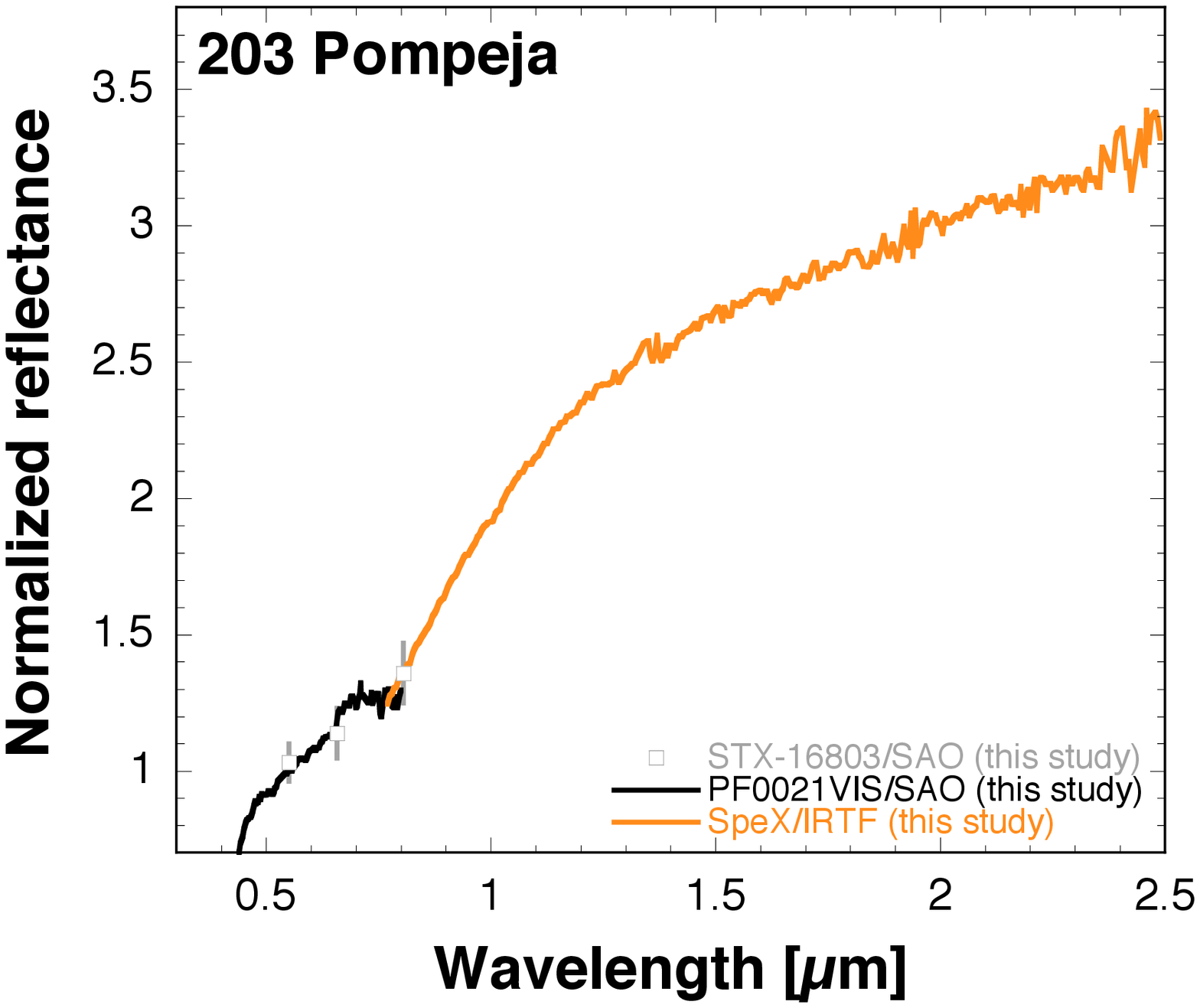}{0.5\textwidth}{}
          \fig{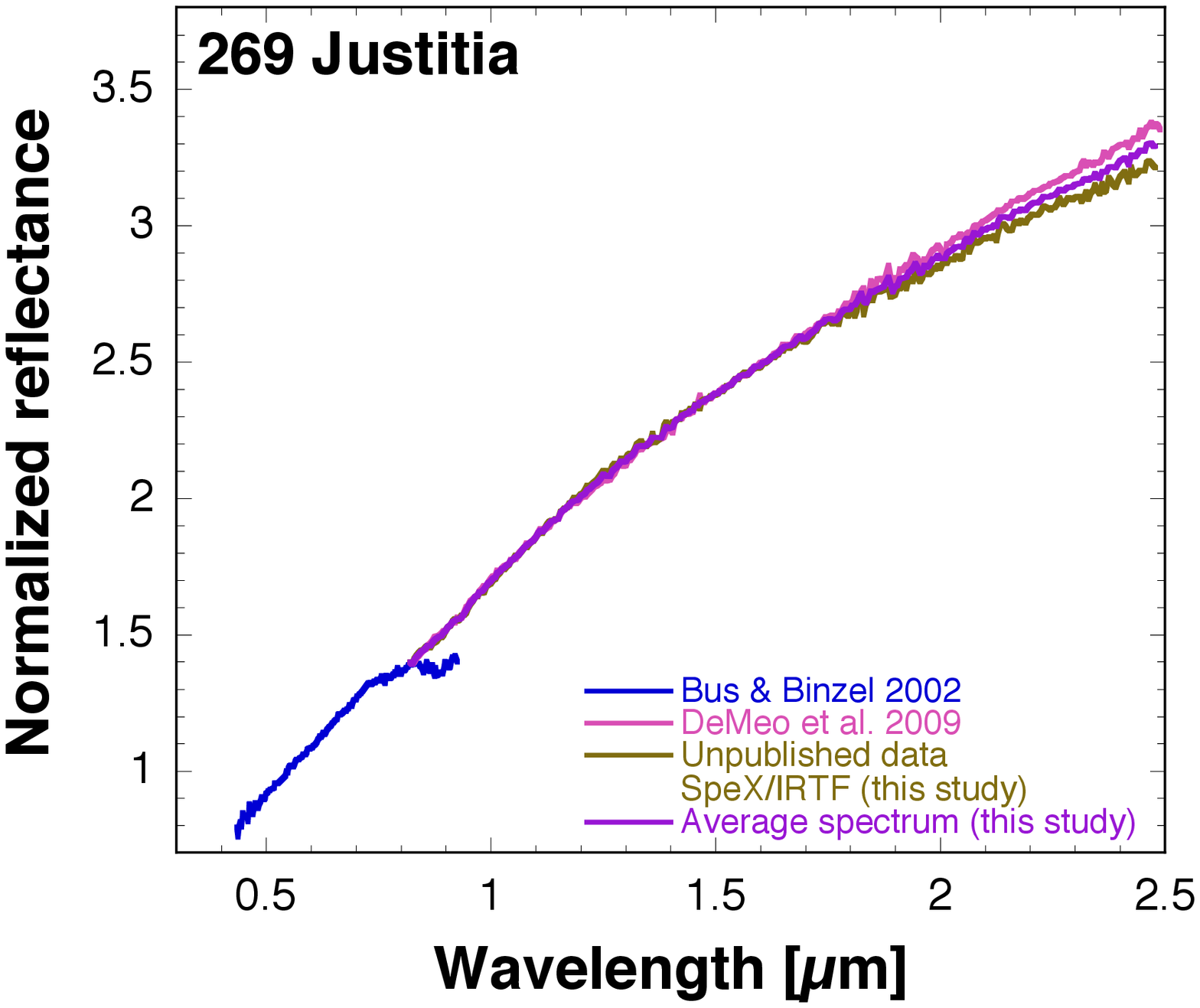}{0.5\textwidth}{}
          }
\caption{
Spectra of 203 Pompeja and 269 Justitia.
}
\label{fig:203-269}
\end{figure*}

\section{Spectral type of the comets, asteroids in cometary orbit, and interstellar objects} \label{sec:listB1}
The taxonomic types of the comet nuclei, asteroids in cometary orbit, and an interstellar object mentioned in this work are listed in Table \ref{tab:B1}.
Objects were classified using the Bus-DeMeo classification method \citep{DeMeo2009} or TNO classification scheme \citep{Fulchignoni2008}.

\section{Polarimetric properties} \label{sec:result2}
Looking through the past literature, we found that polarimetric data for one of the very red asteroids, 269 Justitia, had been obtained \citep{Gil-Hutton2017}.
Therefore, polarization properties of 269 Justitia were compared with those of dark asteroids including a Centaur and a Saturnian satellite (Figure \ref{fig:polarization}).
In terms of polarization properties, the dark objects are divided into the following groups: (1) Ch- and Cgh-asteroids, (2) other dark asteroids including a comet and Jovian Trojans, (3) F-type asteroids, and (4) 269 Justitia, 10199 Chariklo, and Saturn VIII Iapetus on the dark side.
The group to which 269 Justitia belongs has the smallest negative polarization depth and polarization minimum angle.
This indicates that the polarization property of 269 is closer to some icy objects than those of the Jovian Trojans with D-type spectral property.


\begin{figure*}
\setcounter{figure}{0}
\gridline{\fig{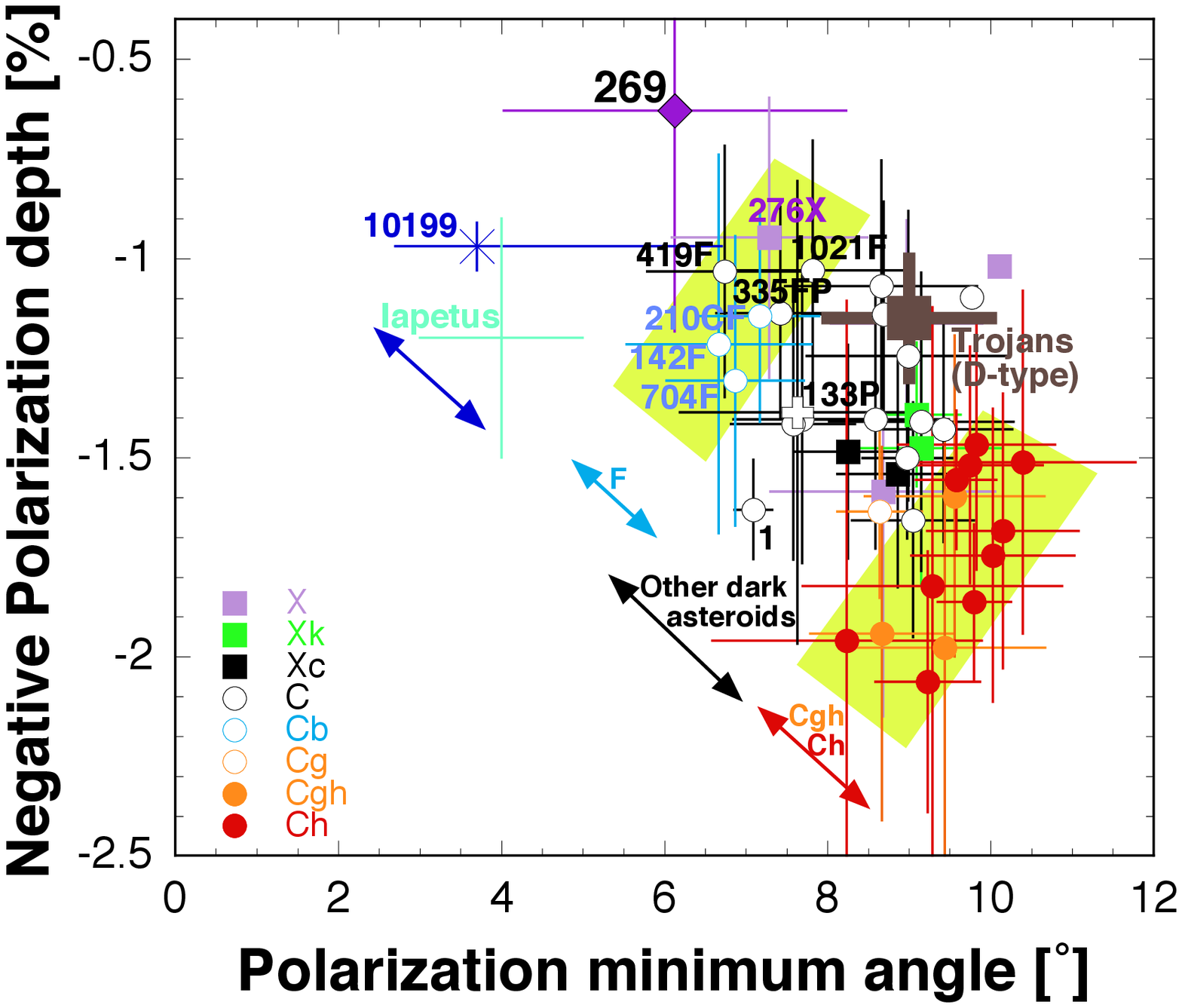}{0.5\textwidth}{}
          }
\caption{Depth of negative polarization versus the phase angle of polarization minimum for dark objects in the Solar System.
The polarimetric data of dark objects are cited from \cite{Gil-Hutton2017}, \cite{Lopez-Sisterna2019}, \cite{Cellino2015}, \cite{Cellino2016}, \cite{Belskaya2010}, \cite{Bagnulo2016}, and \cite{Rosenbush2015}.
The regions of F-type and Ch- or Cgh-type asteroids are shaded in lime green.
269 Justitia inhabits a unique parameter space with a small negative polarization depth and polarization minimum angle.
\label{fig:polarization}
}
\end{figure*}

%
%
%
%
%
%
%
%
\bibliography{hase}
%
%
%
\end{document}